\documentclass[aps,reprint,showpacs,superscriptaddress,groupedaddress]{revtex4-1}
\usepackage{graphicx} 
\usepackage{mathrsfs} 
\usepackage{natbib}
\usepackage{color} 
\begin{document}

\widetext \title{Critical behaviour in the nonlinear elastic response
  of hydrogels}

\author{M.\ Dennison$^{1, 3}$} \author{M.\ Jaspers$^{2}$}
\author{P.H.J.\ Kouwer$^{2}$} \author{C. Storm$^{3}$}
\author{A.E.\ Rowan$^{2}$} \author{F.C.\ MacKintosh$^{1}$}

\affiliation{$^{1}$Department of Physics and Astronomy, Vrije Universiteit, 1081-HV Amsterdam, The Netherlands}

\affiliation{$^{2}$Radboud University Nijmegen, Institute for
  Molecules and Materials, Department of Molecular Materials, 6525-AJ Nijmegen, The Netherlands}

\affiliation{$^{3}$Department of Applied Physics and Institute for
  Complex Molecular Systems, Eindhoven University of Technology, 5600-MB Eindhoven, The Netherlands}

\date{\today}

\begin{abstract}
  In this paper we study the elastic response of synthetic hydrogels
  to an applied shear stress. The hydrogels studied here have
  previously been shown to mimic the behaviour of biopolymer networks
  when they are sufficiently far above the gel point. We show that
  near the gel point they exhibit an elastic response that is
  consistent with the predicted critical behaviour of networks near or
  below the isostatic point of marginal stability. This point
  separates rigid and floppy states, distinguished by the presence or
  absence of finite linear elastic moduli.  Recent theoretical work
  has also focused on the response of such networks to finite or large
  deformations, both near and below the isostatic point.  Despite this
  interest, experimental evidence for the existence of criticality in
  such networks has been lacking. Using computer simulations, we
  identify critical signatures in the mechanical response of
  sub-isostatic networks as a function of applied shear stress.  We
  also present experimental evidence consistent with these
  predictions.  Furthermore, our results show the existence of two
  distinct critical regimes, one of which arises from the nonlinear
  stretch response of semi-flexible polymers.
\end{abstract}

\maketitle

\section{Introduction}
\label{sec:intro}
Highly responsive, or `smart' materials are abundant in Nature;
individual cells, for instance, can adapt their mechanical properties
to the local surroundings by small changes in their internal structure
\cite{janmey2004dealing}. An effective method to enhance the
responsiveness of \emph{synthetic} materials is to operate near a
critical point, where small variations lead to large changes in
material properties. Recent theories have suggested that fibre/polymer
networks can also show critical behaviour near and below the isostatic
point, the point of marginal connectivity separating rigid and floppy
states~\cite{ref:Wyart,ref:ChaseNatPhys2011,ref:MishaPRL2012,ref:Dennison2013,feng2016nonlinear,sharma2016strain}. These
so called marginal networks are predicted to have many interesting and
potentially useful properties that are desirable in smart materials,
being highly sensitive to applied forces and fields and exhibiting an
anomalously high resistance to deformation.

The isostatic point, identified 150 years ago by
Maxwell~\cite{ref:Maxwell1864}, corresponds to the point where the
number of degrees of freedom is just balanced by the number of
constraints imposed by connectivity. This point of marginal stability
has proven to be a rich source of inspiration for novel physics,
ranging from jamming~\cite{ref:PhysRevLett.81.1841,ref:Liu,ref:Hecke}
to zero-temperature critical
behaviour~\cite{ref:Wyart,ref:ChaseNatPhys2011} and even non-quantum
topological matter~\cite{ref:kane2013topological}. However, most of
this work has been theoretical, and experimental realizations of such
criticality have been limited. Granular/colloidal particle packings
show various signatures of criticality, including a shear modulus $G$
that increases continuously with the distance $\phi-\phi_c$ above the
jamming volume fraction
$\phi_c$~\cite{ref:PhysRevLett.81.1841,ref:Liu,ref:Hecke}. Models of
polymer networks and rigidity percolation can show similar critical
behaviour in their linear elastic properties as a function of
connectivity
\cite{ref:LarsonBook,ref:Feng_rigidity,ref:Jacobs_rigidity}.  Jammed
packings and spring/fiber networks are also predicted to exhibit
anomalous stress-strain response near or below the isostatic point,
e.g, with power-law increase of stiffness with stress for systems with
vanishing linear shear
modulus~\cite{ref:Alexander1998,ref:Wyart,ref:SheinmanPRE2012,ref:vanDeen2014,feng2016nonlinear,sharma2016strain}.
Such intrinsically nonlinear elasticity represents a highly responsive
state of matter.

There remain important experimental challenges, however, in creating
fibre or semiflexible polymer networks near a critical point,
including the need to control the connectivity $z$, a key parameter
determining both the isostatic point, as well as the nonlinear
response of sub-isostatic networks to
strain\cite{ref:SheinmanPRE2012,sharma2016strain}. In this paper, we
study hydrogels based on synthetic semi-flexible polymers, (ethylene
glycol)-substituted
polyisocyanides~\cite{ref:nature,ref:jaspers2014ultra}. Dissolved in
water and heated above the gelation temperature (Tgel ~ 19 °C), the
polymers bundle together to form an intertwined network that makes up
the gel. Above their gelation temperature, the network connectivity is
fixed and the materials exhibit a nonlinear elastic response similar
to many biopolymer systems, in which the network stiffness, defined by
the differential shear modulus $K=d\sigma/d\gamma$, increases with
shear stress $\sigma$ as
$\sigma^{3/2}$~\cite{ref:GardelScience2004,ref:LinPRL2010}.  At
temperatures around Tgel, the network morphology is strongly
correlated to the temperature, which allows a high degree of control
over both the network connectivity and the properties of the
individual semi-flexible filaments.

Here, we focus on the regime near and below the gel point, and
demonstrate critical behaviour in the nonlinear stress response of
synthetic hydrogels at low concentrations of order 0.1\% volume
fraction. The networks exhibit a \emph{sub-linear} stiffening response
to an applied shear stress, with $K\propto\sigma^{\alpha<1}$. Using
computer simulations, we show that this unexpected nonlinear elastic
response is a consequence of criticality associated with the isostatic
critical point. Importantly, this work implies that the influence of
isostaticity can extend to network connectivities far below the
isostatic critical point. Furthermore, we find that the intrinsically
nonlinear stretch response of semi-flexible polymer strands in the gel
gives rise to a second anomalous regime, where the networks exhibit a
\emph{super-linear} stiffening response to a shear stress distinct
from the $\sigma^{3/2}$ stiffening commonly associated with
semiflexible polymer networks.

\section{Results}
\label{sec:results}

We have performed rheology experiments and computer simulations in
both linear and nonlinear elastic regimes. Full details of both the
experimental and simulation methods used are given in
Sec.~\ref{sec:method}. In our experimental systems we studied a range
of temperatures near the gel point, while in our simulated networks we
have studied systems at and below the Maxwell isostatic point. We
first present results for the initial stiffening behaviour in both the
experimental and simulated systems, before studying the behaviour at
high shear stress.

\subsection{Low stress regime}

\subsubsection{Experiments}
\label{sec:results_experi1}

\begin{figure}[b]
  \begin{center}
      \includegraphics[height=1.0\columnwidth,angle=270]{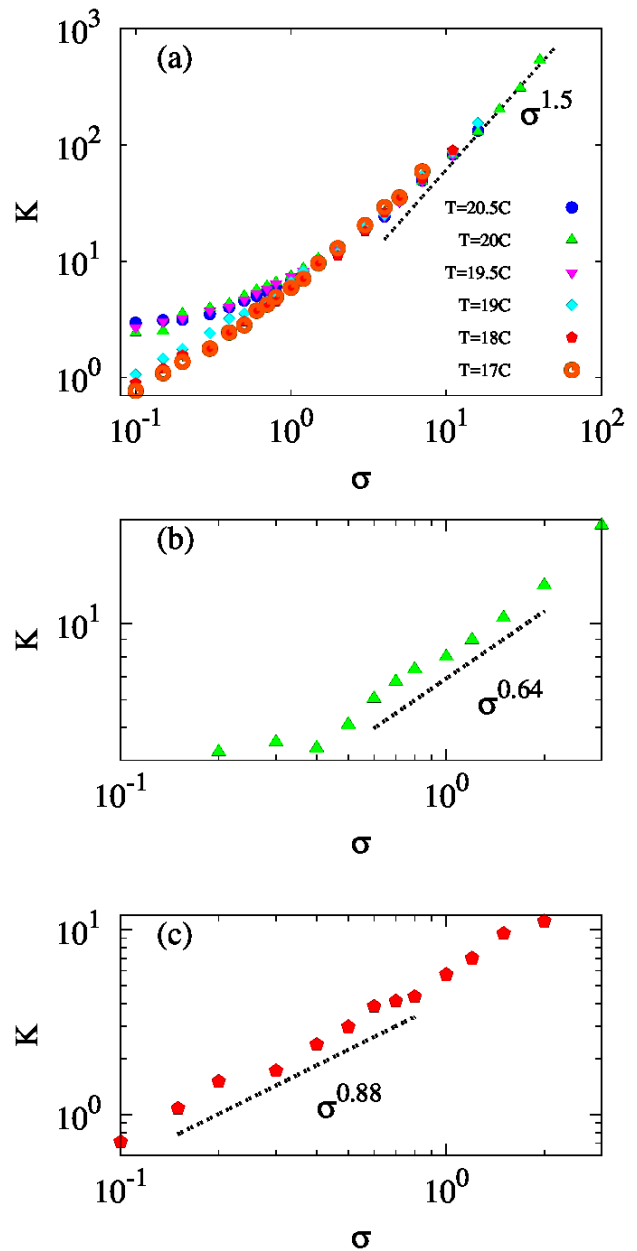}
      \caption{(a) Differential shear modulus $K=
        \partial\sigma/\partial\gamma$, where $\gamma$ is the shear
        strain, against shear stress $\sigma$ for hydrogels at
        temperatures ranging from $T=17^{\circ}$~C to
        $20.5^{\circ}$~C. The lines indicate power-law dependencies of
        $K$ vs $\sigma$. (b) The same data, focusing on the initial
        sublinear scaling regimes for $T=20^{\circ}$~C. (c) The same
        data, focusing on the initial sublinear scaling regimes for
        $T=18^{\circ}$~C. All axes are in units of Pa.}
   \label{fig:experi1}
  \end{center}
\end{figure}

In order to define the network rigidity we have measured the
differential shear modulus, since these hydrogels are incompressible
on experimentally accessible time scales, due to the presence of the
solvent. The differential shear modulus is defined as
$K=\partial\sigma/\partial\gamma$, where $\sigma$ is the shear stress
and $\gamma$ is the shear strain. Cross-linking in our hydrogels
varies with temperature, and well above the gel point
($T\gtrsim30^{\circ}$~C), highly cross-linked gels are formed with a
linear shear moduli
$G_0\simeq100$~Pa~\cite{ref:nature,ref:jaspers2014ultra}. For
$19.5\lesssim T\lesssim21^{\circ}$~C we observe a weak initial linear
elastic regime, followed by a power-law stiffening with
$K\propto\sigma^{\alpha}$ and $\alpha\simeq0.64$ (see
Fig.~\ref{fig:experi1}(b)). The initial $G_0$ vanishes for
temperatures below $T\simeq19^{\circ}$~C, which we identify as the gel
point for our system. For lower temperatures, in the range of $17$ to
$19^{\circ}$~C, we find no apparent linear shear modulus. Instead, we
observe an initial nonlinear regime with $\alpha\simeq0.8$ over about
an order of magnitude in stress, shown in Fig.~\ref{fig:experi1}(c).

\begin{figure}[b]
  \begin{center}
      \includegraphics[width=1.0\columnwidth]{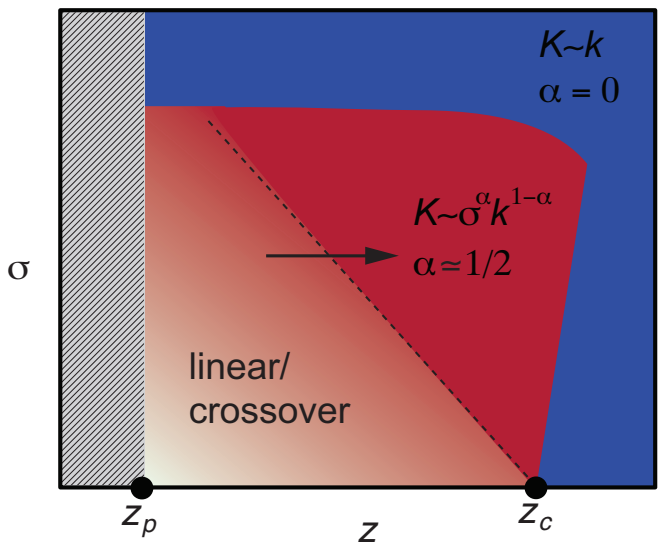}
      \caption{Schematic phase diagram of the various regimes of
        network response as a function of connectivity $z$ (average
        coordination of network nodes) and stress $\sigma$.  Here,
        $z_{c}$ is the critical connectivity, above which a purely
        Hookean network becomes mechanically rigid at zero temperature
        and stress, while $z_{p}$ is the percolation point. The
        exponent $\alpha$ indicates different regimes of stress
        dependence, characterized by network stiffness
        $K\sim\sigma^{\alpha}k^{1-\alpha}_{\mathrm{sp}}$, where $k$ is
        the spring constant. The region labelled `Linear / Cross-over'
        is the regime where we find sub-isostatic networks exhibit an
        initial linear response followed by a cross-over as they enter
        the critical regime.  The critical regime, governed by the
        isostatic point, where $\alpha$ is approximately given by the
        mean-field value of $1/2$, is indicated by the orange central
        triangle. In this region, $\alpha$ is directly related to the
        critical exponents found at $z_c$. Experimentally,
        crosslinking (and therefore $z$) increases with increasing
        temperature, although we are not able to measure $z$ directly
        in our experiments. The expected trend in $z$ with increasing
        temperature is indicated by the arrow.}
   \label{fig:PD}
  \end{center}
\end{figure}

The observed stiffening exponents $\alpha\simeq0.64$ and $0.8$ are
consistent with recently predicted nonlinear elasticity of spring
networks near and below the isostatic, or \emph{marginal},
point~\cite{ref:Wyart,ref:MishaPRL2012,ref:Lubensky_str}. Near
marginal stability, a variety of critical behaviours are predicted,
including a sublinear power law dependence on various stabilizing
fields, such as stress~\cite{ref:Wyart,ref:MishaPRL2012}, thermal
fluctuations \cite{ref:Dennison2013,ref:Lubensky_str} and bending
rigidity \cite{ref:ChaseNatPhys2011}. A simple mean-field argument
suggests the appearance of $K\sim\sigma^{\alpha}$ with critical
stiffening exponent $\alpha\simeq1/2$: in a marginal network, the
linear shear modulus vanishes but any finite stress stabilizes the
network, such that the modulus increases with strain $\gamma$ as
$|\gamma|$, resulting in $\sigma\sim\gamma^2$ and
$d\sigma/d\gamma\sim\sigma^{1/2}$ \cite{ref:Wyart,ref:ChaseReview}.
Thus, an approximate square-root dependence of $K$ on stress is
expected near the critical, or marginal, state, indicated by
$\alpha\simeq1/2$ in the schematic phase diagram in
Fig.\ \ref{fig:PD}. The exponent $\alpha\simeq1/2$ is not, however,
universal. While it is present in triangular-lattice based networks
\cite{ref:ChaseNatPhys2011,ref:MishaPRL2012,ref:Dennison2013} and
random-bond networks \cite{ref:Manon}, for square-lattice based
networks stiffening with an exponent $\alpha\simeq2/3$ has been found
\cite{ref:Lubensky_str}. As we shall show, the critical stiffening of
networks to an applied shear stress can be dependent on the network
topology, with different critical exponents found for different
initial topologies. This indicates that in experimental systems such
as ours, where properties such as the degree of cross-linking and mean
cross-link separation can vary greatly with the temperature, the
critical stiffening exponent may not be universal.

\subsubsection{Simulations}
\label{sec:results_sim1}
In order to understand our experimental observations, we performed
Monte Carlo simulations on both 2D and 3D lattice-based networks. In
our simulation model, we initially use Hookean springs as the model
filaments. Each segment of the filament will resist stretching and
compression with an energy given by
\begin{equation}
\label{eq:E_sp_1}
{\cal U}_{s}=\displaystyle\frac{k}{2}(\ell-\ell_{0})^{2},
\end{equation}
where $k$ is the spring constant, $\ell$ is the length and $\ell_{0}$
is the rest length. Here, in contrast to real semi-flexible polymers,
the filament may stretch indefinitely and they show only a linear
force vs extension, which is not expected to accurately describe our
experimental system at high stress, but can describe the behaviour at
low stress. The filament stiffness is controlled by the spring
constant $k$, which is related to the ratio of the persistence length
to the segment contour length $\ell_{p}/\ell_{0}$, see
Eq.~(\ref{eq:lp}). Furthermore, we also include a bending energy
between sequential segments along a single filament, given by
\begin{equation}
\label{eq:E_sp_bend_1}
{\cal U}_{b}=\frac{\kappa}{2} \theta^{2}_{ij},
\end{equation}
where $\kappa$ is the bending coefficient and $\theta_{ij}$ is the
angle between segements $i$ and $j$.

\begin{figure}[!h]
  \begin{center}
      \includegraphics[height=1.0\columnwidth,angle=270]{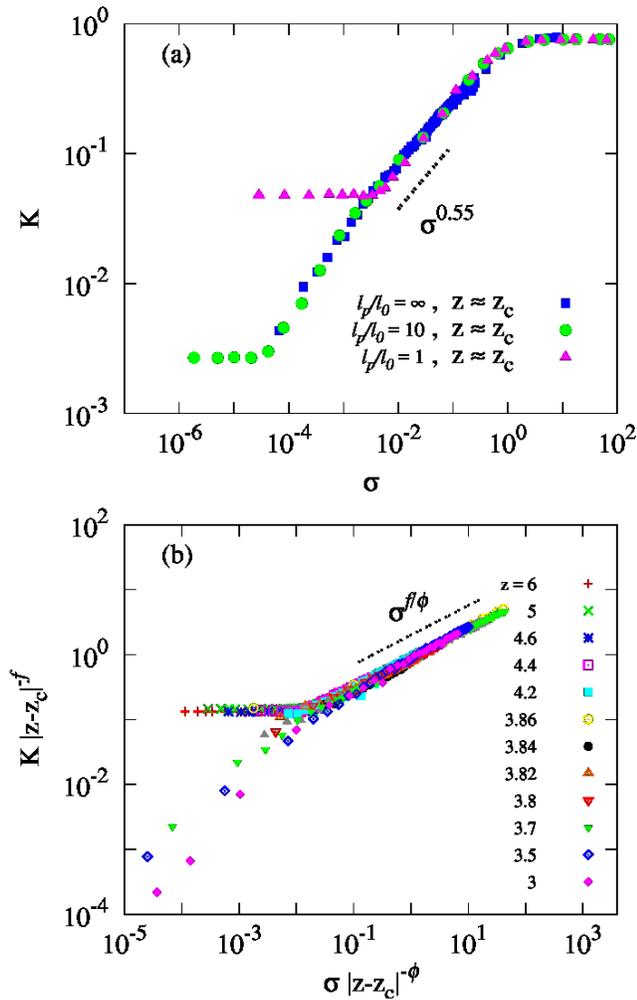}
      \caption{(a) Differential shear modulus $K$ vs stress $\sigma$
        for 2D Hookean spring networks with $z=3.85\alt
        z_{c}\simeq3.857$. $K$ and $\sigma$ are in units of the spring
        constant $k$. Squares show $\ell_{p}/\ell_{0}=\infty$
        (athermal), circles $\ell_{p}/\ell_{0}=10$ and triangles
        $\ell_{p}/\ell_{0}=1$ where $\ell_{p}$ is the persistence
        length. We estimate that our experimental systems are in the
        range $\ell_{p}/\ell_{0}=1-10$
        \cite{ref:nature,ref:jaspers2014ultra}. Dashed line shows
        $K\sim\sigma^{\alpha}$ dependence, and indicates the region
        over which we fit to find $\alpha$.  (b) Scaling collapse of
        the differential shear modulus $K$, as a function of the shear
        stress $\sigma$ and the distance $\Delta z = z-z_{\mathrm{c}}$
        from the critical connectivity, using the scaling anstatz
        given in Eq.~\ref{eq:widom}. Data shown is for 2D Hookean
        spring networks with $\ell_{p}/\ell_{0}=10$, and with
        connectivities in the range $3.0-6.0$. Here, $f=1.4\pm0.03$
        and $\phi=2.6\pm0.1$.}
   \label{fig:sim}
  \end{center}
\end{figure}

\begin{figure}[!h]
  \begin{center}
      \includegraphics[height=1.0\columnwidth,angle=270]{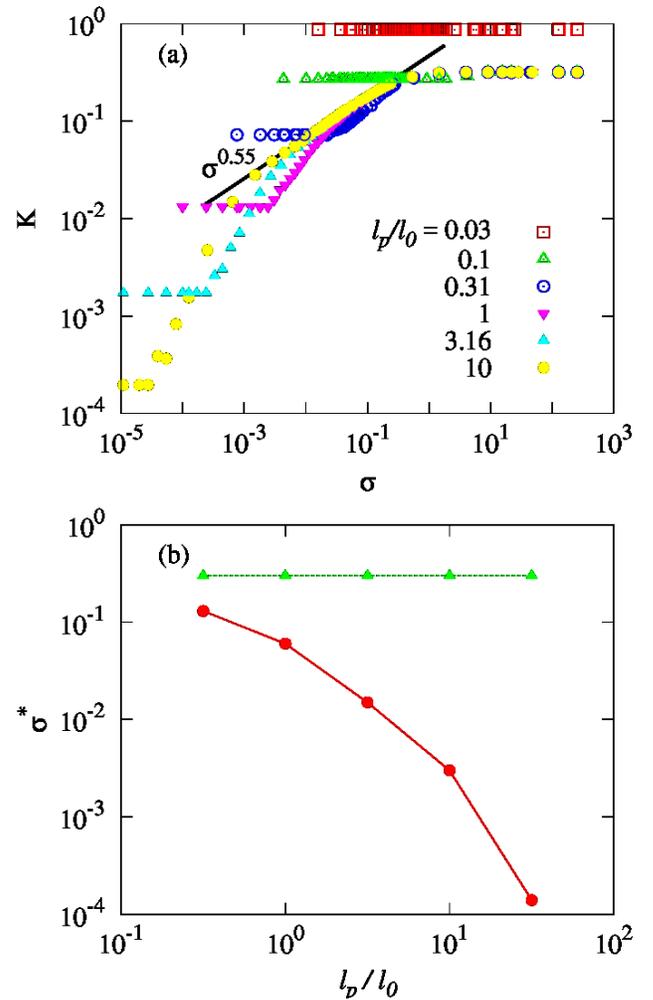}
      \caption{(a) Differential shear modulus $K$ vs
          stress $\sigma$ for 2D Hookean spring networks with
          $z=3$. $K$ and $\sigma$ are in units of the spring constant
          $k$. Legend indicates $\ell_{p}/\ell_{0}$, where $\ell_{p}$
          is the persistence length. We estimate that our experimental
          systems are in the range $\ell_{p}/\ell_{0}=1-10$
          \cite{ref:nature,ref:jaspers2014ultra}. Dashed line shows
          $K\sim\sigma^{\alpha}$ dependence in the critical
          regime. (b) Stress $\sigma^{*}$ at which the above networks
          enter (red circles) and leave (green triangles) the critical
          regime, where we observe a $K\sim\sigma^{0.55}$ dependence,
          for networks with a range of $\ell_{p}/\ell_{0}$ values.}
   \label{fig:sim_z3d0}
  \end{center}
\end{figure}

\begin{figure}[!h]
  \begin{center}
      \includegraphics[height=1.0\columnwidth,angle=270]{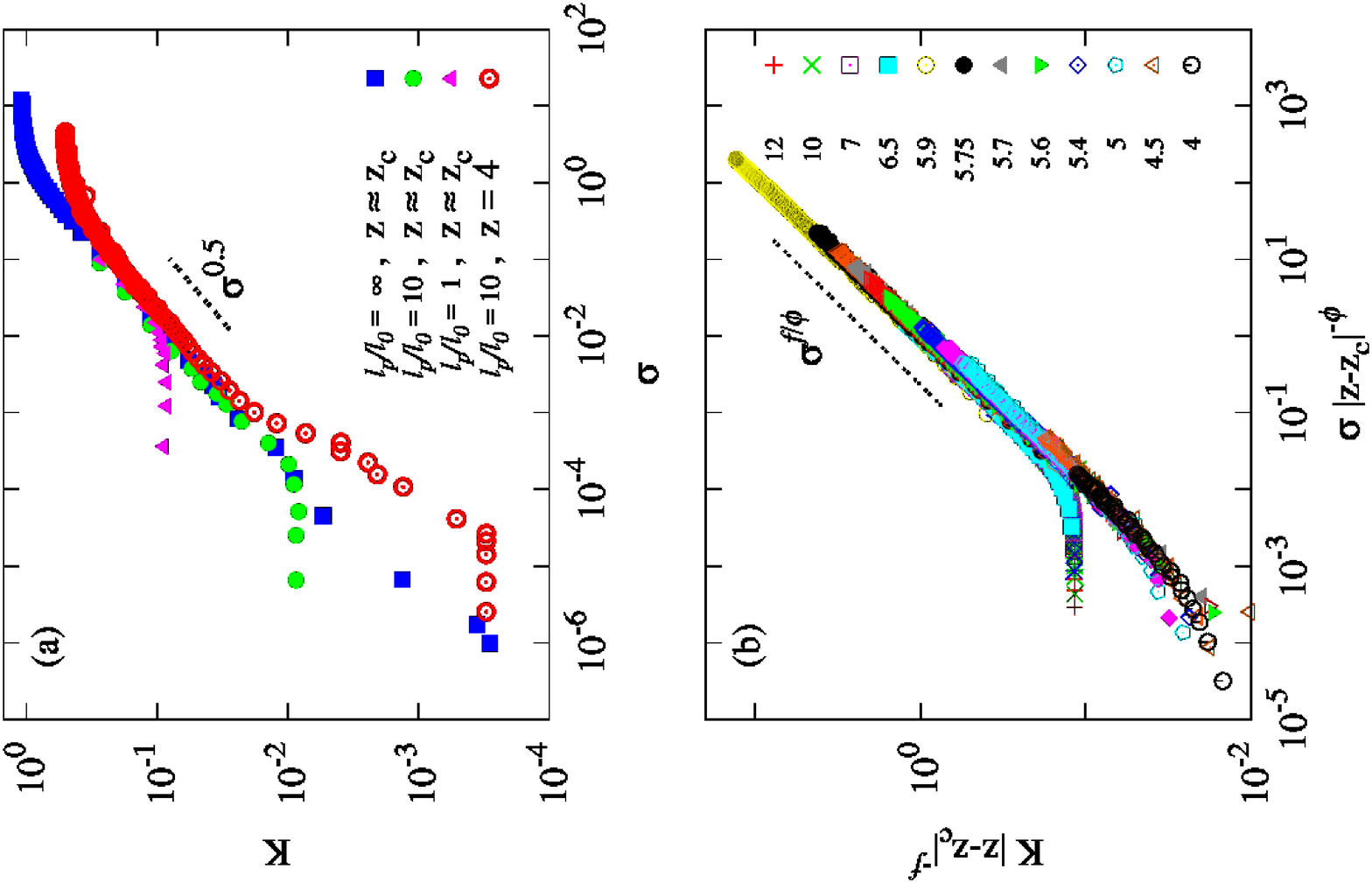}
      \caption{(a) Main plot: Differential shear modulus $K$ vs stress
        $\sigma$ for 3D Hookean spring networks with $z=5.84\alt
        z_{c}\simeq5.844$ (solid symbols) and $z=4$ (open
        symbols). $K$ and $\sigma$ are in units of
        $k/\ell_{0}$, where $k$ is the
        spring constant and $\ell_{0}$ is the rest length of the
        springs (which also gives the segment contour length). Squares
        show $\ell_{p}/\ell_{0}=\infty$ (athermal), circles
        $\ell_{p}/\ell_{0}=10$ and triangles $\ell_{p}/\ell_{0}=1$
        where $\ell_{p}$ is the persistence length. We estimate that
        our experimental systems are in the range
        $\ell_{p}/\ell_{0}=1-10$ \cite{ref:nature,ref:jaspers2014ultra}. Dashed lines show
        $K\sim\sigma^{\alpha}$ dependencies, and indicate the region
        over which we fit to find $\alpha$. Inset: $K$ vs $\sigma$ for
        3D networks with $z=5.84\alt z_{c}\simeq5.844$ using
        semi-flexible filaments (see methods), with
        $\ell_{p}/\ell_{0}=10$. Both axes are in units of
        $k/\ell_{0}$, where here $k$ is an
        effective spring constant. (b) Scaling collapse
        of the differential shear modulus $K$, as a function of the
        shear stress $\sigma$ and the distance $\Delta z =
        z-z_{\mathrm{c}}$ from the critical connectivity, using the
        scaling anstatz given in Eq.~\ref{eq:widom}. Data shown is for
        3D Hookean spring networks with $\ell_{p}/\ell_{0}=10$, and
        with connectivities in the range $4.0-12.0$. Here,
        $f=1.6\pm0.1$ and $\phi=3.2\pm0.1$.}
      \label{fig:sim_3d}
  \end{center}
\end{figure}

For systems at the marginal point (Fig.~\ref{fig:sim}(a) and
Fig.~\ref{fig:sim_3d}(a)), we find that athermal networks
($\ell_{p}/\ell_{0}\rightarrow\infty$ or $k\rightarrow\infty$) exhibit
no initial linear response to an applied shear strain $\gamma$. As
$\ell_{p}/\ell_{0}$ (and hence $k$) decreases, thermal fluctuations
give rise to a linear response regime for both marginal and
submarginal networks, in which we find $K=G_{0}$. We note that this
initial linear shear modulus $G_{0}$ can depend on either the thermal
or the bend energy, depending on which of the two energy scales
dominates. In the former case $G_{0}$ behaves as in
Ref.~\cite{ref:Dennison2013}, while in the later it would behave as in
Ref.~\cite{ref:ChaseNatPhys2011}. The results shown here are for
$\kappa=0$, although in practice we find that both the temperature and
the bending rigidity only affect this initial linear regime, and not
the subsequent stiffening behaviour. This linear regime is followed by
an increase in $K$ once the stress exceeds a threshold $\sigma_{0}$,
giving a $K\sim\sigma^{\alpha}$ dependence with $\alpha<1$. For 2D
networks we observe a stiffening exponent at the marginal point of
$\alpha\sim0.55\pm0.02$, found by fitting over the region indicated in
Fig~\ref{fig:sim}(a), while for 3D networks we find
$\alpha\sim0.5\pm0.02$, indicated in Fig~\ref{fig:sim_3d}(a). As the
modulus and stress have the same units, on dimensional grounds $K$
should show an additional dependence on another energy scale, which we
find to be the spring constant, scaling as
$K\sim\sigma^{\alpha}k^{1-\alpha}_{\mathrm{sp}}$ in 2D and
$K\sim\sigma^{\alpha}(k/\ell_{0})^{1-\alpha}_{\mathrm{sp}}$ in
3D. Finally, at high stresses, we see that $K$ becomes invariant to
$\sigma$ and begins to scale as $K\sim k$, corresponding to pure
stretching of the springs.

Below the marginal state, in an initially floppy regime, the
dependence of $K$ on the stabilizing field $\sigma$ is less
clear. Prior work has shown that the \emph{critical} regime with
$\alpha\simeq1/2$ is not limited to systems finely-tuned to the
na\"ive isostatic connectivity $z_c$, but also extends to much lower
connectivities when the networks are stabilized by other interactions,
such as
stress~\cite{ref:Dennison2013,ref:MishaPRL2012,sharma2016strain}. Indeed,
we find that even submarginal networks exhibit the observed critical
stiffening behaviour, as can be seen in Fig~\ref{fig:sim_z3d0}(a) and
Fig~\ref{fig:sim_3d}(a), where networks with a connectivity well below
$z_{c}$ are taken into a regime where they stiffen as
$K\sim\sigma^{\alpha}$ as stress is increased. Thus our experimental
networks, which we expect to be submarginal with connectivity
$z\lesssim4$ ($z_{c}\sim6$ in $3D$), would be taken into a critical
regime by an applied shear stress, where the network should show
sublinear stiffening. We note that the size of the
$K\sim\sigma^{\alpha}$ stiffening regime is sensitive to the ratio
$\ell_{p}/\ell_{0}$; if this ratio is too small, $G_{0}$ will be large
enough to dominate the response, as can seen in
Fig~\ref{fig:sim_z3d0}(a). In
  Fig~\ref{fig:sim_z3d0}(b) we plot the stress at which sub-marginal
  2D networks (with $z=3$) enter the critical regime as a function of
  the ratio of persistence length to segment length. As can be seen,
  networks with a higher value of $\ell_{p}/\ell_{0}$ (corresponding
  to stiff filaments) will enter the critical regime at a much lower
  stress than networks with a lower value.

In order to examine if this is true critical behaviour we have
calculated the non-affine fluctuations of the system, which are known to
diverge at critical points in elastic networks. We first define the
differential non-affinity $\Gamma$ as
\begin{equation}
\label{eq:non_aff}
\Gamma = \displaystyle\frac{1}{\ell_{0}^{2}}\displaystyle\frac{\displaystyle\left\langle\Delta{\bf r}^{2}\right\rangle}{\left(\Delta\gamma\right)^{2}},
\end{equation}
where ${\bf r}={\bf r}_{na}-{\bf r}_{a}$ is the non-affine
contribution to the node displacement, with ${\bf r}_{na}$ the
position of a node and ${\bf r}_{a}$ the position if the displacement
would have been affine. $\langle\dots\rangle$ denotes the average over
all nodes. A high value of $\Gamma$ means that the network deformation
is more differentially non-affine, while a low value means it is less
so.

\begin{figure}[!h]
  \begin{center}
      \includegraphics[height=0.85\columnwidth,angle=270]{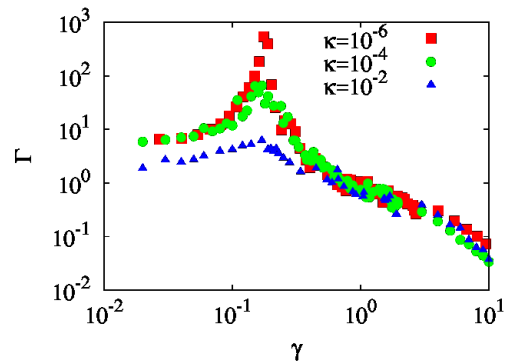}
      \caption{Differential non-affinity $\Gamma$ against applied
        shear strain $\gamma$ for 2D Hookean spring networks with
        connectivity $z=3.5$. Data is for various bending rigidities
        $\kappa$.}
\label{fig:NA}
  \end{center}
\end{figure}

Figure~\ref{fig:NA} shows $\Gamma$ against the applied shear strain
$\gamma$ for networks with connectivity $z=3.5$ (well below the
marginal point $z_{c}=3.857$), simulated at a range of bending
rigidities $\kappa$ (see Eq.~(\ref{eq:E_sp_bend})) in the athermal
limit of $\ell_{p}/\ell_{0}\rightarrow\infty$. We choose to plot our
data against $\gamma$ instead of $\sigma$ as the networks will enter
the critical regime at similar strains but vastly different
stresses. For low bending rigidities $\kappa$ we find that, as the
strain is increased, $\Gamma$ increases, reaching a peak at a value
corresponding to the network entering the critical regime, where the
differential shear modulus scales as $K\sim\sigma^{\alpha}$. Beyond
this peak $\Gamma$ decreases with increasing $\gamma$. This divergence
of the differential non-affinity is further evidence that we are in a
true critical regime. For higher values of $\kappa$ we find that the
peak value decreases, until eventually no divergence is found,
indicating that in this case the bending rigidity suppresses
criticality, consistent with previous work
\cite{ref:ChaseNatPhys2011}.

In the critical regime, the stiffening exponent $\alpha=f/\phi$ is
directly related to critical exponents $f$ and $\phi$ defined by
\begin{equation}
\label{eq:widom}
K = k|\Delta z|^{f} \mathscr{F}\left(\frac{\sigma}{k}|\Delta z|^{-\phi}\right),
\end{equation}
which is demonstrated in Fig.~\ref{fig:sim}(b) and
Fig.~\ref{fig:sim_3d}(b). Such cross-over scaling has been
demonstrated previously for the critical point of random resistor networks
\cite{ref:Straley}, fiber networks \cite{ref:ChaseNatPhys2011} and both
athermal and thermal spring networks
\cite{ref:Wyart,ref:Dennison2013}. Here we find $f=1.4\pm0.03$ and
$\phi=2.6\pm0.1$ (with $\alpha=0.54$) for 2D networks and $f=1.6$ and
$\phi=3.2$ (with $\alpha=0.5$) for 3D networks. These stiffening
exponents are comparable to the values observed experimentally and
close to the mean-field value of $\alpha=1/2$
\cite{ref:Wyart}. However, as noted previously, different network
topologies can exhibit different critical exponents. Random bond
networks have been shown to exhibit mean-field like stiffening with
temperature $T$ at the critical point, with $G_{0}\sim T^{\alpha}$
where $\alpha=f/\phi=1/2$ with $f=1$ and $\phi=2$. Square lattices,
which are marginal objects, stiffen with $G_{0}\sim T^{\alpha}$ where
$\alpha\simeq2/3$ \cite{ref:Lubensky_str}.

\begin{figure}[!h]
  \begin{center}
      \includegraphics[height=1.0\columnwidth,angle=270]{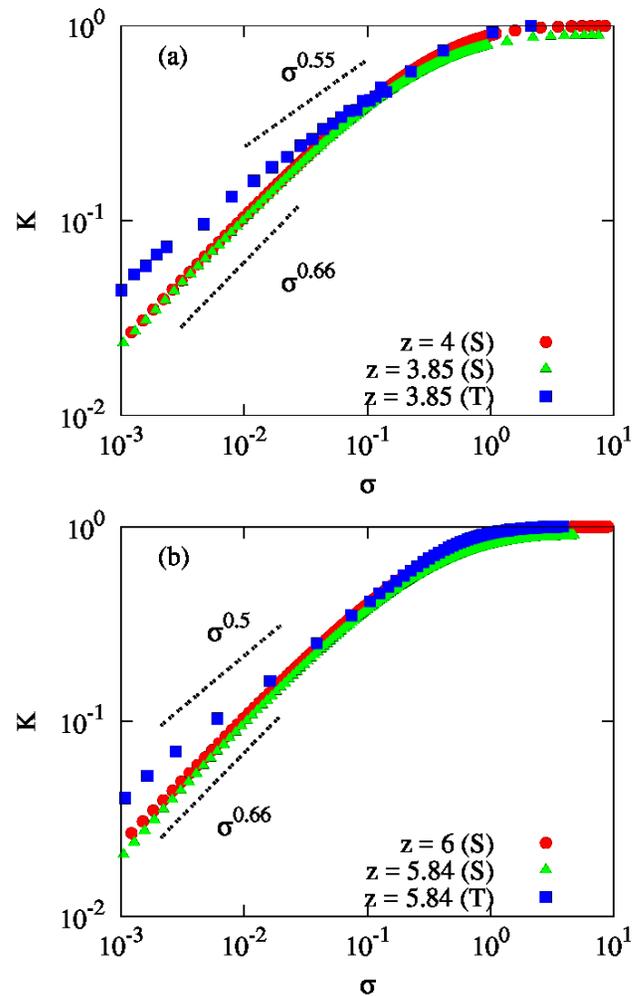}
      \caption{(a) Differential shear modulus $K$ vs stress $\sigma$
        for 2D Hookean spring networks using triangular lattice based
        networks at the marginal point $z=3.85\simeq z_{c}$ (squares)
        and square lattice networks with $z=4$ (circles) and $z=3.85$
        (triangles). $K$ and $\sigma$ are in units of the spring
        constant $k$. Dashed lines show
        $K\sim\sigma^{\alpha}$ dependencies, and indicate the region
        over which we fit to find $\alpha$. Data for the triangular
        lattice system has been shifted to better show the observed
        stiffening. (b) $K$ against $\sigma$ for 3D Hookean spring
        networks using FCC lattice based networks at the marginal
        point $z=5.84\simeq z_{c}$ (squares) and simple cubic lattice
        networks with $z=6$ (circles) and $z=5.84$ (triangles). $K$
        and $\sigma$ are in units of
        $k/\ell_{0}$. Dashed lines show
        $K\sim\sigma^{\alpha}$ dependencies, and indicate the region
        over which we fit to find $\alpha$. Data for the FCC lattice
        system has been shifted to better show the observed
        stiffening.}
   \label{fig:sim_sq}
  \end{center}
\end{figure}

In order to see how systems with different topologies stiffen with an
applied shear stress we have simulated a square lattice network in 2D
and a simple cubic lattice in 3D, and the results are shown in
Fig.~\ref{fig:sim_sq}(a), where we compare the stiffening to that of
triangular (2D) and FCC (3D) lattice based networks. These networks
are marginal objects, as any deformation will result in a cost in
energy, and as can be seen, the network stiffens as
$K\sim\sigma^{\alpha}$ with $\alpha\sim0.66$, distinct from the
$\alpha\sim0.55$ found for the triangular lattice network. The same
behaviour can be seen for 3d networks using a simple cubic lattice,
and exhibits stiffening with $\alpha\sim0.66$
(Fig.~\ref{fig:sim_sq}(b)), again distinct from the $\alpha\sim0.5$
found for the FCC lattice network. This implies that the starting
topology is important: two networks can stiffen with two different
critical exponents depending on the initial topology, with the network
with a higher possible local $z$ ($z=12$ for an FCC lattice) have a
lower critical stiffening exponent than that with a lower possible
local $z$ ($z=6$ for a simple cubic lattice).

\subsection{High stress regime}

\subsubsection{Experiments}
\label{sec:results_experi2}

\begin{figure}[!h]
  \begin{center}
      \includegraphics[height=1.0\columnwidth,angle=270]{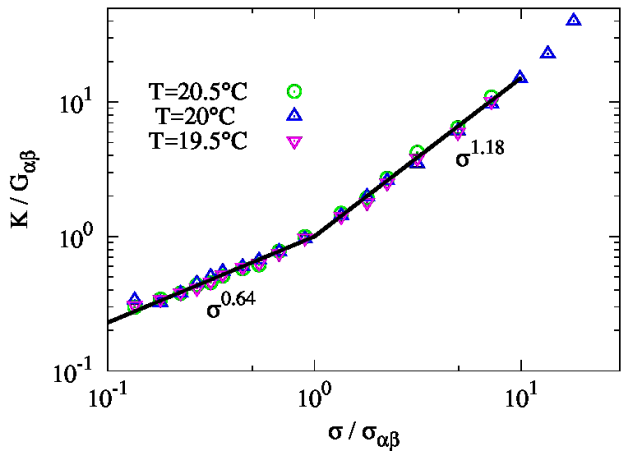}
      \caption{Differential shear modulus $K$ against stress
        $\sigma$ measured at temperatures from $T=19.5$ to
        $20.5^{\circ}$~C. $K$ and $\sigma$ are normalized by the
        empirical values $G_{\alpha\beta}$ and $\sigma_{\alpha\beta}$,
        the differential modulus and shear stress at the crossover
        from the $\alpha$ to $\beta$ regimes. Solid lines show
        $K\sim\sigma^{0.64}$ and $K\sim\sigma^{1.18}$
        dependence.}
   \label{fig:experi2}
  \end{center}
\end{figure}

As the stress in our experimental systems is increased beyond the
initial $K\sim\sigma^{\alpha}$ regime, we find a second stiffening
regime, for which we define a second exponent $\beta\simeq1.2$, as can
be seen in Fig.~\ref{fig:experi1} and more clearly in
Fig.~\ref{fig:experi2}. This exponent is distinct from the asymptotic
exponent of ${1.5}$ observed previously for this system
\cite{ref:nature}. We hypothesize that this arises from the nonlinear
spring constant of the polymers making up the network
\cite{ref:MacKintosh1995,ref:GardelScience2004,ref:Storm2005,ref:LinPRL2010,ref:nature}.
As the stress increases, these polymers stretch and enter a nonlinear
regime characterized by a force-extension relation in which the force
\begin{equation}
f\sim1/|1-\epsilon|^2\label{FK-divergence}
\end{equation}
depends on the
relative extension $\epsilon$ \cite{ref:Fixman1973,ref:Marko1995,ref:MacKintosh1995} . This leads to an effective spring
constant $k_\mathrm{sp}\propto f^{3/2}$, where $f\propto\sigma$ is the
force on the segment. At low stresses the force-extension relation is
linear $f\sim\epsilon$, and hence the effective spring constant is
independent of stress. As we have shown from our simulation results,
the network stiffness scales as $K\sim\sigma^\alpha\times
k^{1-\alpha}$ \cite{ref:MishaPRL2012}. Thus, if we
substitute in $k_\mathrm{sp}\propto \sigma^{3/2}$, we predict an
initial $K\sim\sigma^\alpha$ regime at low stresses, followed by
$K\sim\sigma^\beta$ at higher stresses, where
\begin{equation}
\label{eq:beta}
\beta=3/2-\alpha/2.
\end{equation}

Beyond this regime, at very high stresses, we see evidence of the
$K\sim\sigma^{1.5}$ regime expected for semi-flexible polymer
networks.

\subsubsection{Simulations}
\label{sec:results_sim2}

\begin{figure}[!h]
  \begin{center}
      \includegraphics[height=1.0\columnwidth,angle=270]{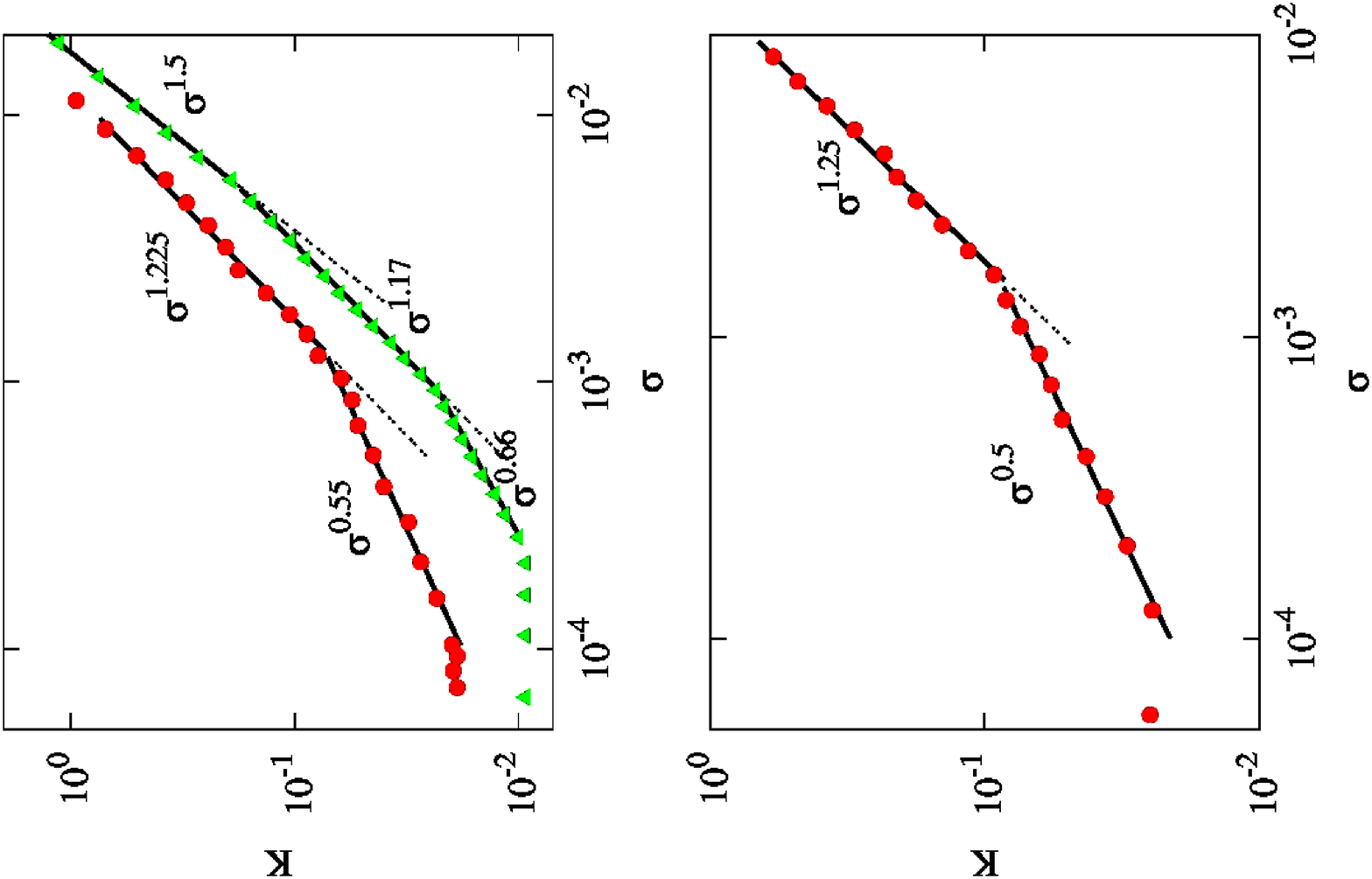}
      \caption{(a)Differential shear modulus $K$ vs stress $\sigma$
        for 2D networks with $z=3.85\alt z_{c}\simeq3.857$ using
        semi-flexible filaments (see methods), with
        $\ell_{p}/\ell_{0}=10$, indicated by the red circles. Both
        axes are in units of $k$, where here $k$ is an effective
        spring constant, see Eq.~(\ref{eq:lp}). Lines show
        $K\sim\sigma^{\alpha}$ and $K\sim\sigma^{\beta}$ dependencies,
        and indicate the region over which we fit to find $\alpha$ and
        $\beta$. Green triangles indicate data for a diluted square
        lattice network, also with $z=3.857$, with the
        $K\sim\sigma^{\alpha}$, $K\sim\sigma^{\beta}$ and
        $K\sim\sigma^{1.5}$ dependencies indicated. (b) $K$ vs
        $\sigma$ for 3D networks with $z=5.84\alt z_{c}\simeq5.844$
        using semi-flexible filaments (see methods), with
        $\ell_{p}/\ell_{0}=10$. Both axes are in units of
        $k/\ell_{0}$.}
   \label{fig:sim_nl}
  \end{center}
\end{figure}

In order to understand this intermediate regime in our experimental
system, corresponding to $1$Pa $\alt\sigma\alt10$Pa in
Fig.~\ref{fig:experi1} where we observed $K\sim\sigma^{\beta}$, we
simulated networks using the nonlinear response of semi-flexible
filaments. Full details are given in the methods section,
Eq.~(\ref{eq:E_sp2}) to Eq.~(\ref{eq:E_bend}). Here the filament
stretch response is initially linear, followed by a strong stiffening
due to the pulling out of thermal bending modes. As for the Hookean
spring model, the filament stiffness is controlled by the effective
spring constant $k$, related to $\ell_{p}/\ell_{0}$, see
Eq.~(\ref{eq:lp}). Our results are shown for 2D in
Fig.~\ref{fig:sim_nl}(a) and 3D networks in
Fig.~\ref{fig:sim_nl}(b). Following an initial regime with
$K\sim\sigma^{\alpha}$, we find both $K\sim\sigma^{\beta}$ and
$K\sim\sigma^{3/2}$ regimes, where $\beta$ obeys
Eq.~(\ref{eq:beta}). We stress that this relation holds for both 2D
and 3D networks, as well as for different initial network
topologies. This can be seen in Fig.~\ref{fig:sim_nl}(a), where we
also show data for diluted square lattice networks, which show
$\alpha\sim0.66$ (as in Fig.~\ref{fig:sim_sq}), and $\beta\sim1.17$.
These results are also consistent with the phase diagram in Fig.\
\ref{fig:PD}, with both $K\sim\sigma^\alpha\times
k^{1-\alpha}_{\mathrm{sp}}$ at intermediate stress and $K\sim k$ at
high stress, where $k\propto\sigma^{3/2}$, consistent with known
extensional properties of semi-flexible
polymers~\cite{ref:MacKintosh1995,ref:GardelScience2004,ref:Storm2005,ref:LinPRL2010,ref:nature,ref:Fixman1973,ref:Marko1995,ref:ChaseReview}.
These observations can account for our experimental results for $1$Pa
$\alt\sigma\alt10$Pa.

\begin{figure}[!h]
  \begin{center}
      \includegraphics[height=1.0\columnwidth,angle=270]{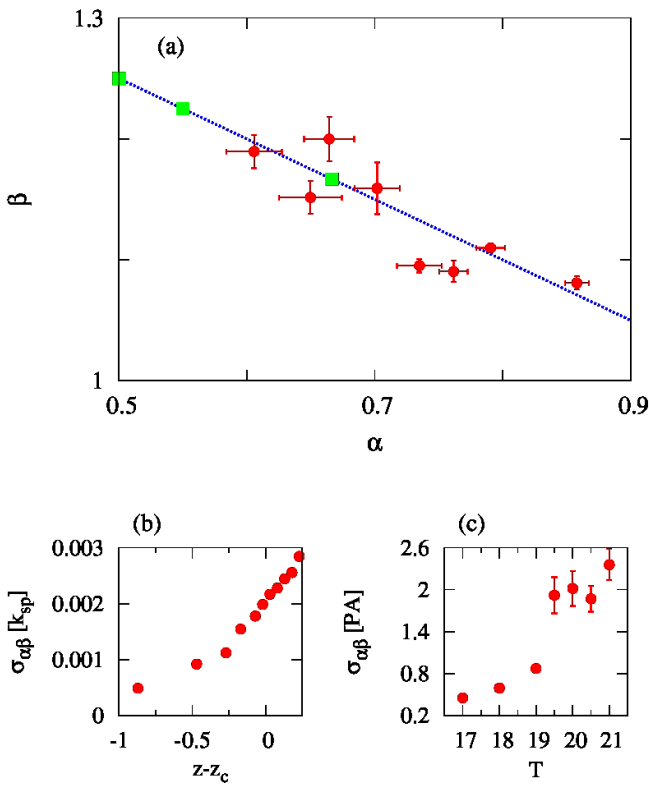}
      \caption{(a) Values of $\alpha$ and $\beta$ found for individual
        temperatures ranging from $T=17$ to $21^{\circ}$~C, indicated
        by the red circles. Line shows the relation
        $\beta=1.5-0.5\alpha$, while green squares show simulation
        data for 2d (triangular lattice with $\alpha\sim0.55$ and
        square lattice with $\alpha \sim 0.66$) and 3d
        ($\alpha\sim0.5$) networks.  (b) Stress $\sigma_{\alpha\beta}$
        at the cross-over from the initial sublinear $\alpha$
        stiffening regime to the second, superlinear $\beta$ regime,
        against distance from the critical connectivity $z-z_{c}$ for
        simulated 3D networks using the potential in
        Eqs.~(\ref{eq:E_sp2})-(\ref{eq:E_bend}) to model the
        filaments. Here $\sigma_{\alpha\beta}$ is in units of the
        effective spring constant $k$ given in Eq.~(\ref{eq:ksp}). (c)
        $\sigma_{\alpha\beta}$ (given in $Pa$) for experimental
        networks, against temperature $T$. }
   \label{fig:experi3}
  \end{center}
\end{figure}

In Fig.~\ref{fig:experi2}, we show data for experimental networks at
$T=19.5-20.5^{\circ}$~C with $K$ (and $\sigma$) scaled by the shear
modulus $G_{\alpha\beta}$ (and stress $\sigma_{\alpha\beta}$) at which
we observe the cross-over from $\alpha$ to $\beta$ regimes. Here we
find excellent agreement of the data using an initial stiffening
exponent $\alpha\sim0.64$, followed by $\beta\sim1.18$ obtained from
Eq.~(\ref{eq:beta}). In Fig.~\ref{fig:experi3}(a) we show the values
of $\alpha$ and $\beta$ found for individual temperatures from $17$ to
$21^{\circ}$~C, which again show good agreement with the relation
given in Eq.~(\ref{eq:beta}). Importantly, the prediction (dashed
line) contains no adjustable parameters. Finally, we also observe an
evolution of the cross-over stress $\sigma_{\alpha\beta}$, which
increases with increasing $T$ (Fig.~\ref{fig:experi3}(c)). This is
also consistent with the predicted trend with increasing $z$, shown in
Fig.~\ref{fig:experi3}(b). This behaviour further indicates that the
connectivity increases with the temperature, as indicated in the
schematic phase diagram in Fig.~\ref{fig:PD} by the solid arrow,
corresponding to the variation of $z$ one would expect for increasing
temperature.

\section{Conclusions}
\label{sec:Conclusions}

We have shown that synthetic hydrogels exhibit an elastic response
consistent with predicted critical behaviour associated with isostatic
and subisostatic networks. Our results show that these hydrogels
stiffen sublinearly under an applied shear stress and exhibit a critical
stiffening exponent that is consistent with that predicted theoretically
and in computer simulations. Perhaps surprisingly, even networks deep into the
subisostatic regime are predicted to exhibit such behaviour. This can account for
our experimental results, where all the samples are expected to be in the sub-isostatic regime, 
corresponding to local connectivities $z\lesssim4<z_c$ in 3D. 
This suggests that the
marginal point is an important control mechanism for the elastic
stability of networks in response to an applied shear stress, and our
results support the proposed phase diagram in Fig.~\ref{fig:PD},
indicating a very broad range over which critical control of mechanics
is possible. Furthermore, both our simulation and experimental results
show the existence of a second regime of critical stiffening, where
the individual filaments exhibit a non-linear response to stretch
deformation resulting in a superlinear stiffening regime.

This work also identifies the ratio of the persistence length to the
cross-link separation, $\ell_{p}/\ell_{0}$, as key design parameters:
this should be of the order of $0.1-10$ to achieve critical control of
network mechanics.  When the polymers are too flexible, for instance,
the linear shear modulus dominates the stiffening behaviour. Thus, for
synthetic polymers that are usually flexible, controlled bundle
formation may be important for future materials development using
these principles. This work demonstrates an experimentally realizable
system that exhibits mechanical critical behaviour, opening the way
for further experimental studies of marginal/isostatic networks.

\begin{acknowledgments}
  We acknowledge financial support from FOM/NWO (M.D, C.S, F.C.M),
  NRSCC (M.J, A.E.R) NWO Gravitation (A.E.R, P.H.J.K) and NanoNextNL
  (A.E.R, P.H.J.K).  We would like to thank David Weitz for fruitful
  discussions and suggestions.\\

  M.D, C.S and F.C.M designed the simulations. M.D. performed the
  simulations. M.J, P.H.J.K and A.E.R designed the experimental
  work. M.J. synthesised the polymers and carried out the mechanical
  tests. All authors contributed to the writing of the paper.

\end{acknowledgments}

\section{Methods}
\label{sec:method}

\subsection{Experiments}
\label{sec:method_experi}

\begin{figure}[!h]
  \begin{center}
      \includegraphics[width=0.75\columnwidth]{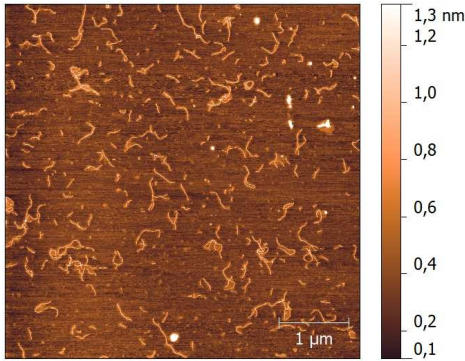}
    \caption{AFM image of individual polymers spincoated from a
      $0.1$~mg/mL solution of the polymer in dichloromethane on
      freshly cleaved Mica. The scale bar is indicated in the figure.}
\label{fig:AFM}
  \end{center}
\end{figure}

\begin{figure}[!h]
  \begin{center}
      \includegraphics[height=0.75\columnwidth,angle=270]{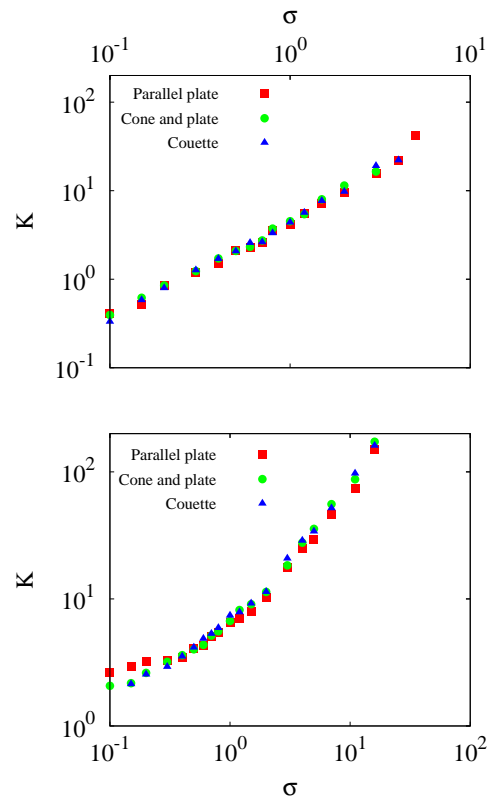}
    \caption{Differential modulus $K$ as a function of stress $\sigma$
      for three different geometries at (a) $T=17^{\circ}$~C (in the
      `pre-gel' regime) and (b) at $T=20^{\circ}$~C (just above the
      gel point). All axes are in units of Pa.}
\label{fig:exp}
  \end{center}
\end{figure}

Our gels were synthesized and purified following a previously
described procedure \cite{ref:nature}, and an AFM image of individual
polymers is shown in Fig.~\ref{fig:AFM}. The catalyst/monomer ratio of
$1:2000$ yielded a polymer of average molecular weight $M_{v} =$
400~kg~mol$^{-1}$ as determined by viscometry. For gel studies, the
polymer was dissolved in purified water (milliQ) by stirring for at
least $24$~hrs at $4$~$^{\circ}$C.  Rheology was performed, by
default, with a stress-controlled rheometer (Discovery HR-1, TA
Instruments) with an aluminium parallel plate geometry ($40$~mm
diameter) and a gap of $500~\mathrm{\mu m}$. Samples were inserted in
the rheometer at $5$~$^{\circ}$C (i.e. as a liquid) and gelation
occurred between the setup by raising the temperature using a peltier
plate. Drying of the sample was prevented by maintaining a moist
atmosphere. The storage modulus in the linear regime was obtained by
applying an oscillatory strain of $1\%$ at a frequency of $1$~$Hz$ and
measuring the sinusoidal stress response. The non-linear regime was
probed by applying a steady pre-stress $\sigma$ to the sample and
superposing a small oscillatory stress with an amplitude of
$|\delta\sigma| < 0.1 \sigma$ at a frequencies $0.1-10$~$Hz$. The
differential modulus was calculated from the oscillatory strain
response $\delta\gamma$, as $K =
\partial\sigma/\partial\gamma=\delta\sigma/\delta\gamma$.  To
investigate the role of the non-linear strain-field that of the
parallel plate setup, we measured for selected samples the mechanical
properties in a Couette and cone and plate geometry
(Fig.~\ref{fig:exp}). The experimental details are as follows: Cone
and plate geometry: aluminium, $40$~mm diameter, cone angle
$1$~$^{\circ}$, truncation gap $29~\mathrm{\mu m}$; Couette geometry:
aluminium cup and bob, cup diameter $30.41$~mm, bob diameter
$27.98$~mm (thus a gap of $1.22$~mm), bob length $42.10$~mm.

Although the strain field in the plate-plate geometry is not constant,
the linear and non-linear mechanical properties are well
represented. As an illustration, we show the (non-)linear
mechanical properties at two different temperatures in three different
configurations: plate-plate; cone-plate and Couette, see
Fig.~\ref{fig:exp} For both temperatures, the results from either
configuration are similar over the entire stress range, in line for
what was observed in gels based on actin \cite{ref:GardelScience2004}.

\subsection{Simulations}
\label{sec:method_sim}

We use the Monte Carlo method to simulate two- and three-dimensional
lattice-based networks. The network nodes are initially arranged on a
triangular lattice in 2D and on an FCC lattice in 3D, and are allowed
to fluctuate off-lattice during the simulations. Nearest neighbour
nodes are connected with model filament segments to give a fully
connected network with $z=6$ in 2D and $z=12$ in 3D, before random
segments are removed to lower the network connectivity $z$.

In this work have used two types of model filaments. The first is the
Hookean spring model, which has been used in many previous
studies~\cite{ref:ChaseNatPhys2011,ref:MishaPRL2012,ref:Chase_length,ref:Dennison2013},
where the stretching or compression of a filament segment $i$ will
involve a cost in energy given by
\begin{equation}
\label{eq:E_sp}
{\cal U}_{s}=\displaystyle\frac{k}{2}(\ell_{i}-\ell_{0,i})^{2},
\end{equation}
where $\ell_{i}$ is the length and $\ell_{0,i}$ is the contour length
of segment $i$, and $k$ is the spring constant. In such a model the
segments may stretch indefinitely and show only a linear
force-extension for any deformation. A bending rigidity is also
incorporated, where the energy cost for bending two coaxially
connected springs, $i$ and $j$, is given by
\begin{equation}
  \label{eq:E_sp_bend}
  {\cal U}_{b}=\frac{\kappa}{2} \theta^{2}_{ij},
\end{equation}
where $\kappa$ is the bending rigidity and $\theta_{ij}$ is the angle
between springs $i$ and $j$.

In order to capture the non-linear response of semi-flexible polymers
to stretching and compression we include a nonlinear spring to represent the known force-extension 
appropriate for stiff chain segments of length $\ell_0\lesssim\ell_p$ \cite{odijk1995stiff,ref:MacKintosh1995,ref:Storm2005,ref:ChaseReview},
which can be well-approximated by a combination of the divergence in Eq.~\ref{FK-divergence}, together with a linear spring \cite{ref:GardelScience2004,ref:HuismanPRE,ref:ChaseReview}. Here, we follow the approach presented in
Ref.~\cite{ref:HuismanPRE} and use a nonlinear potential to more accurately
describe the experimental filament response \cite{ref:single}. This can be
summarized by the force along a stretched segment which is given by
\begin{equation}
\label{eq:E_sp2}
f=\frac{9k_\mathrm{B}T\ell_p}{\ell_{0,i}^2}\left[\frac{1}{(1-\epsilon)^2}-1-\frac{1}{3}\epsilon\right],
\end{equation}
where $T$ is the temperature, $k_{\mathrm{b}}$ the Boltzmann constant
and $\epsilon_{i}$ is the scaled extension of filament $i$, given by
\begin{equation}
\label{eq:g_sp}
\epsilon_{i} = \frac{1}{6} + \frac{\ell_{p}\ell_{i}}{\ell^{2}_{0,i}} - \frac{\ell_{p}}{\ell_{0,i}},
\end{equation}
where $\ell_{p}$ is the persistence length of the filaments. For
lattice networks the contour lengths are identical for all
segments. The energy due to a deformation of filament segment $i$ is
then given by
\begin{equation}
\label{eq:E_sp2}
\frac{{\cal U}_{s}}{k_{\mathrm{b}}T}=\left\{
\begin{array}{l l} 
    \displaystyle -\frac{9\epsilon^{2}_{i}[5+6\epsilon_{i}]}{6\epsilon_{i}-1} \;\; &\epsilon_{i} > 0,
   \\ \displaystyle \left|\pi^{2}\epsilon_{i}-\frac{\pi^{4}}{90}(\exp[90\epsilon_{i}/\pi^{2}]-1)\right|\;\; &\epsilon_{i} < 0,
\end{array}\right.
\end{equation}
The top line in Eq.~(\ref{eq:E_sp2}) gives the energy cost for
stretching, while the bottom line that for compression. In this more
realistic model a bending energy is also applied to the system, which
again acts on pairs of filament segments that are connected coaxially
at the network nodes. The energy due to the bending of connected
filaments $i$ and $j$ is given by \cite{ref:HuismanPRE}
\begin{equation}
\label{eq:E_bend}
\frac{{\cal U}_{b}}{k_{\mathrm{b}}T}=
\frac{\ell_{p}\theta^{2}_{ij}}{\ell_{0,i} + \ell_{0,j}},
\end{equation}
where $\theta_{ij}$ is the angle between the end-to-end vectors of
filament segments $i$ and $j$. 

The stiffness of Hookean springs is controlled by the spring constant
$k$, while for the semi-flexible potential it is
controlled by the persistence length $\ell_{p}$. These can be related
through an effective spring constant given by
\begin{equation}
\label{eq:ksp}
k=
\frac{90k_{\mathrm{b}}T\ell^{2}_{p}}{\ell^{4}_{0}},
\end{equation}
or alternatively by the reduced persistence length
\begin{equation}
\label{eq:lp}
\ell^{\prime}_{p}=\frac{\ell_{p}}{\ell_{0}}=\displaystyle\sqrt{\frac{k\ell^{2}_{0}}{90k_{\mathrm{b}}T}}
\end{equation}
We note that for small deformations of a segment the energy cost is
the same for both the semi-flexible and the Hookean spring potentials
when the spring constants (or reduced persistence lengths) are equal.

The networks are then sheared using Lees-Edwards boundary conditions
\cite{ref:Lees_Edwards}. After applying a shear strain $\gamma$ to the
system, we calculate the shear stress $\sigma$ and differential
modulus $K$ as described in Refs.~\cite{ref:Hoover,ref:Dennison2013}.

\bibliography{main}

\providecommand*{\mcitethebibliography}{\thebibliography}
\csname @ifundefined\endcsname{endmcitethebibliography}
{\let\endmcitethebibliography\endthebibliography}{}
\begin{mcitethebibliography}{36}
\providecommand*{\natexlab}[1]{#1}
\providecommand*{\mciteSetBstSublistMode}[1]{}
\providecommand*{\mciteSetBstMaxWidthForm}[2]{}
\providecommand*{\mciteBstWouldAddEndPuncttrue}
  {\def\EndOfBibitem{\unskip.}}
\providecommand*{\mciteBstWouldAddEndPunctfalse}
  {\let\EndOfBibitem\relax}
\providecommand*{\mciteSetBstMidEndSepPunct}[3]{}
\providecommand*{\mciteSetBstSublistLabelBeginEnd}[3]{}
\providecommand*{\EndOfBibitem}{}
\mciteSetBstSublistMode{f}
\mciteSetBstMaxWidthForm{subitem}
{(\emph{\alph{mcitesubitemcount}})}
\mciteSetBstSublistLabelBeginEnd{\mcitemaxwidthsubitemform\space}
{\relax}{\relax}

\bibitem[Janmey and Weitz(2004)]{janmey2004dealing}
P.~A. Janmey and D.~A. Weitz, \emph{Trends in biochemical sciences}, 2004,
  \textbf{29}, 364--370\relax
\mciteBstWouldAddEndPuncttrue
\mciteSetBstMidEndSepPunct{\mcitedefaultmidpunct}
{\mcitedefaultendpunct}{\mcitedefaultseppunct}\relax
\EndOfBibitem
\bibitem[Wyart \emph{et~al.}(2008)Wyart, Liang, Kabla, and
  Mahadevan]{ref:Wyart}
M.~Wyart, H.~Liang, A.~Kabla and L.~Mahadevan, \emph{Phys. Rev. Lett.}, 2008,
  \textbf{101}, 215501\relax
\mciteBstWouldAddEndPuncttrue
\mciteSetBstMidEndSepPunct{\mcitedefaultmidpunct}
{\mcitedefaultendpunct}{\mcitedefaultseppunct}\relax
\EndOfBibitem
\bibitem[Broedersz \emph{et~al.}(2011)Broedersz, Lubensky, Mao, and
  MacKintosh]{ref:ChaseNatPhys2011}
C.~P. Broedersz, T.~C. Lubensky, X.~Mao and F.~C. MacKintosh, \emph{Nature
  Physics}, 2011, \textbf{7}, 983\relax
\mciteBstWouldAddEndPuncttrue
\mciteSetBstMidEndSepPunct{\mcitedefaultmidpunct}
{\mcitedefaultendpunct}{\mcitedefaultseppunct}\relax
\EndOfBibitem
\bibitem[Sheinman \emph{et~al.}(2012)Sheinman, Broedersz, and
  MacKintosh]{ref:MishaPRL2012}
M.~Sheinman, C.~P. Broedersz and F.~C. MacKintosh, \emph{Phys. Rev. Lett.},
  2012, \textbf{109}, 238101\relax
\mciteBstWouldAddEndPuncttrue
\mciteSetBstMidEndSepPunct{\mcitedefaultmidpunct}
{\mcitedefaultendpunct}{\mcitedefaultseppunct}\relax
\EndOfBibitem
\bibitem[Dennison \emph{et~al.}(2013)Dennison, Sheinman, Storm, and
  MacKintosh]{ref:Dennison2013}
M.~Dennison, M.~Sheinman, C.~Storm and F.~C. MacKintosh, \emph{Phys. Rev.
  Lett.}, 2013, \textbf{111}, 095503\relax
\mciteBstWouldAddEndPuncttrue
\mciteSetBstMidEndSepPunct{\mcitedefaultmidpunct}
{\mcitedefaultendpunct}{\mcitedefaultseppunct}\relax
\EndOfBibitem
\bibitem[Feng \emph{et~al.}()Feng, Levine, Mao, and Sander]{feng2016nonlinear}
J.~Feng, H.~Levine, X.~Mao and L.~M. Sander, \emph{Soft Matter\kern-1mm},
  DOI:10.1039/C5SM01856K (2016)\relax
\mciteBstWouldAddEndPuncttrue
\mciteSetBstMidEndSepPunct{\mcitedefaultmidpunct}
{\mcitedefaultendpunct}{\mcitedefaultseppunct}\relax
\EndOfBibitem
\bibitem[Sharma \emph{et~al.}()Sharma, Licup, Jansen, Rens, Sheinman,
  Koenderink, and MacKintosh]{sharma2016strain}
A.~Sharma, A.~J. Licup, K.~A. Jansen, R.~Rens, M.~Sheinman, G.~H. Koenderink
  and F.~C. MacKintosh, \emph{Nature Physics\kern-1mm},  DOI: 10.1038/NPHYS3628
  (2016)\relax
\mciteBstWouldAddEndPuncttrue
\mciteSetBstMidEndSepPunct{\mcitedefaultmidpunct}
{\mcitedefaultendpunct}{\mcitedefaultseppunct}\relax
\EndOfBibitem
\bibitem[Maxwell(1864)]{ref:Maxwell1864}
J.~C. Maxwell, \emph{Philos. Mag.}, 1864, \textbf{27}, 297\relax
\mciteBstWouldAddEndPuncttrue
\mciteSetBstMidEndSepPunct{\mcitedefaultmidpunct}
{\mcitedefaultendpunct}{\mcitedefaultseppunct}\relax
\EndOfBibitem
\bibitem[Cates \emph{et~al.}()Cates, Wittmer, Bouchaud, and
  Claudin]{ref:PhysRevLett.81.1841}
M.~E. Cates, J.~P. Wittmer, J.-P. Bouchaud and P.~Claudin, \emph{Phys. Rev.
  Lett.}, \textbf{81}, 1841\relax
\mciteBstWouldAddEndPuncttrue
\mciteSetBstMidEndSepPunct{\mcitedefaultmidpunct}
{\mcitedefaultendpunct}{\mcitedefaultseppunct}\relax
\EndOfBibitem
\bibitem[Liu and Nagel(2010)]{ref:Liu}
A.~J. Liu and S.~R. Nagel, \emph{Annu. Rev. Condens. Matter Phys.}, 2010,
  \textbf{1}, 347\relax
\mciteBstWouldAddEndPuncttrue
\mciteSetBstMidEndSepPunct{\mcitedefaultmidpunct}
{\mcitedefaultendpunct}{\mcitedefaultseppunct}\relax
\EndOfBibitem
\bibitem[van Hecke(2010)]{ref:Hecke}
M.~van Hecke, \emph{J. Phys.: Condens. Matter}, 2010, \textbf{22}, 033101\relax
\mciteBstWouldAddEndPuncttrue
\mciteSetBstMidEndSepPunct{\mcitedefaultmidpunct}
{\mcitedefaultendpunct}{\mcitedefaultseppunct}\relax
\EndOfBibitem
\bibitem[Kane and Lubensky(2013)]{ref:kane2013topological}
C.~L. Kane and T.~C. Lubensky, \emph{Nature Physics}, 2013, \textbf{10},
  39\relax
\mciteBstWouldAddEndPuncttrue
\mciteSetBstMidEndSepPunct{\mcitedefaultmidpunct}
{\mcitedefaultendpunct}{\mcitedefaultseppunct}\relax
\EndOfBibitem
\bibitem[Larson(1998)]{ref:LarsonBook}
R.~G. Larson, \emph{The Structure and Rheology of Complex Fluids}, Oxford
  University Press, 1998\relax
\mciteBstWouldAddEndPuncttrue
\mciteSetBstMidEndSepPunct{\mcitedefaultmidpunct}
{\mcitedefaultendpunct}{\mcitedefaultseppunct}\relax
\EndOfBibitem
\bibitem[Feng and Sen(1984)]{ref:Feng_rigidity}
S.~Feng and P.~N. Sen, \emph{Phys. Rev. Lett.}, 1984, \textbf{52}, 216\relax
\mciteBstWouldAddEndPuncttrue
\mciteSetBstMidEndSepPunct{\mcitedefaultmidpunct}
{\mcitedefaultendpunct}{\mcitedefaultseppunct}\relax
\EndOfBibitem
\bibitem[Jacobs and Thorpe(1996)]{ref:Jacobs_rigidity}
D.~J. Jacobs and M.~F. Thorpe, \emph{Phys. Rev. E}, 1996, \textbf{53},
  3682\relax
\mciteBstWouldAddEndPuncttrue
\mciteSetBstMidEndSepPunct{\mcitedefaultmidpunct}
{\mcitedefaultendpunct}{\mcitedefaultseppunct}\relax
\EndOfBibitem
\bibitem[Alexander(1998)]{ref:Alexander1998}
S.~Alexander, \emph{Phys. Rep.}, 1998, \textbf{296}, 65\relax
\mciteBstWouldAddEndPuncttrue
\mciteSetBstMidEndSepPunct{\mcitedefaultmidpunct}
{\mcitedefaultendpunct}{\mcitedefaultseppunct}\relax
\EndOfBibitem
\bibitem[Sheinman \emph{et~al.}(2012)Sheinman, Broedersz, and
  MacKintosh]{ref:SheinmanPRE2012}
M.~Sheinman, C.~P. Broedersz and F.~C. MacKintosh, \emph{Phys. Rev. E}, 2012,
  \textbf{85}, 021801\relax
\mciteBstWouldAddEndPuncttrue
\mciteSetBstMidEndSepPunct{\mcitedefaultmidpunct}
{\mcitedefaultendpunct}{\mcitedefaultseppunct}\relax
\EndOfBibitem
\bibitem[van Deen \emph{et~al.}(2014)van Deen, Simon, Zeravcic, Dagois-Bohy,
  Tighe, and van Hecke]{ref:vanDeen2014}
M.~S. van Deen, J.~Simon, Z.~Zeravcic, S.~Dagois-Bohy, B.~P. Tighe and M.~van
  Hecke, \emph{Phys. Rev. E}, 2014, \textbf{90}, 020202\relax
\mciteBstWouldAddEndPuncttrue
\mciteSetBstMidEndSepPunct{\mcitedefaultmidpunct}
{\mcitedefaultendpunct}{\mcitedefaultseppunct}\relax
\EndOfBibitem
\bibitem[Kouwer \emph{et~al.}(2013)Kouwer, Koepf, Le~Sage, Jaspers, van Buul,
  Eksteen-Akeroyd, Woltinge, Schwartz, Kitto, Hoogenboom, Picken, Nolte,
  Mendes, and Rowan]{ref:nature}
P.~H.~J. Kouwer, M.~Koepf, V.~A.~A. Le~Sage, M.~Jaspers, A.~M. van Buul, Z.~H.
  Eksteen-Akeroyd, T.~Woltinge, E.~Schwartz, H.~J. Kitto, R.~Hoogenboom, S.~J.
  Picken, R.~J.~M. Nolte, E.~Mendes and A.~E. Rowan, \emph{Nature}, 2013,
  \textbf{493}, 651\relax
\mciteBstWouldAddEndPuncttrue
\mciteSetBstMidEndSepPunct{\mcitedefaultmidpunct}
{\mcitedefaultendpunct}{\mcitedefaultseppunct}\relax
\EndOfBibitem
\bibitem[Jaspers \emph{et~al.}(2014)Jaspers, Dennison, Mabesoone, MacKintosh,
  Rowan, and Kouwer]{ref:jaspers2014ultra}
M.~Jaspers, M.~Dennison, M.~F. Mabesoone, F.~C. MacKintosh, A.~E. Rowan and
  P.~H. Kouwer, \emph{Nature communications}, 2014, \textbf{5}, year\relax
\mciteBstWouldAddEndPuncttrue
\mciteSetBstMidEndSepPunct{\mcitedefaultmidpunct}
{\mcitedefaultendpunct}{\mcitedefaultseppunct}\relax
\EndOfBibitem
\bibitem[Gardel \emph{et~al.}(2004)Gardel, Shin, MacKintosh, Mahadevan,
  Matsudaira, and Weitz]{ref:GardelScience2004}
M.~L. Gardel, J.~H. Shin, F.~C. MacKintosh, L.~Mahadevan, P.~Matsudaira and
  D.~A. Weitz, \emph{Science}, 2004, \textbf{304}, 1301\relax
\mciteBstWouldAddEndPuncttrue
\mciteSetBstMidEndSepPunct{\mcitedefaultmidpunct}
{\mcitedefaultendpunct}{\mcitedefaultseppunct}\relax
\EndOfBibitem
\bibitem[Lin \emph{et~al.}(2010)Lin, Yao, Broedersz, Herrmann, MacKintosh, and
  Weitz]{ref:LinPRL2010}
Y.~C. Lin, N.~Y. Yao, C.~P. Broedersz, H.~Herrmann, F.~C. MacKintosh and D.~A.
  Weitz, \emph{Phys. Rev. Lett}, 2010, \textbf{104}, 058101\relax
\mciteBstWouldAddEndPuncttrue
\mciteSetBstMidEndSepPunct{\mcitedefaultmidpunct}
{\mcitedefaultendpunct}{\mcitedefaultseppunct}\relax
\EndOfBibitem
\bibitem[Mao \emph{et~al.}(2015)Mao, Souslov, Mendoza, and
  Lubensky]{ref:Lubensky_str}
X.~Mao, A.~Souslov, C.~I. Mendoza and T.~C. Lubensky, \emph{Nature
  Communications}, 2015, \textbf{6}, 5968\relax
\mciteBstWouldAddEndPuncttrue
\mciteSetBstMidEndSepPunct{\mcitedefaultmidpunct}
{\mcitedefaultendpunct}{\mcitedefaultseppunct}\relax
\EndOfBibitem
\bibitem[Broedersz and MacKintosh(2014)]{ref:ChaseReview}
C.~P. Broedersz and F.~C. MacKintosh, \emph{Rev. Mod. Phys.}, 2014,
  \textbf{86}, 995\relax
\mciteBstWouldAddEndPuncttrue
\mciteSetBstMidEndSepPunct{\mcitedefaultmidpunct}
{\mcitedefaultendpunct}{\mcitedefaultseppunct}\relax
\EndOfBibitem
\bibitem[Wigbers \emph{et~al.}(2015)Wigbers, MacKintosh, and
  Dennison]{ref:Manon}
M.~C. Wigbers, F.~C. MacKintosh and M.~Dennison, \emph{Phys. Rev. E}, 2015,
  \textbf{92}, 042145\relax
\mciteBstWouldAddEndPuncttrue
\mciteSetBstMidEndSepPunct{\mcitedefaultmidpunct}
{\mcitedefaultendpunct}{\mcitedefaultseppunct}\relax
\EndOfBibitem
\bibitem[Straley(1976)]{ref:Straley}
J.~Straley, \emph{J. Phys. C: Solid State Phys.}, 1976, \textbf{9}, 783\relax
\mciteBstWouldAddEndPuncttrue
\mciteSetBstMidEndSepPunct{\mcitedefaultmidpunct}
{\mcitedefaultendpunct}{\mcitedefaultseppunct}\relax
\EndOfBibitem
\bibitem[MacKintosh \emph{et~al.}(1995)MacKintosh, Kas, and
  Janmey]{ref:MacKintosh1995}
F.~C. MacKintosh, J.~Kas and P.~Janmey, \emph{Phys. Rev. Lett.}, 1995,
  \textbf{75}, 4425\relax
\mciteBstWouldAddEndPuncttrue
\mciteSetBstMidEndSepPunct{\mcitedefaultmidpunct}
{\mcitedefaultendpunct}{\mcitedefaultseppunct}\relax
\EndOfBibitem
\bibitem[Storm \emph{et~al.}(2005)Storm, Pastore, MacKintosh, Lubensky, and
  Janmey]{ref:Storm2005}
C.~Storm, J.~J. Pastore, F.~C. MacKintosh, T.~C. Lubensky and P.~A. Janmey,
  \emph{Nature}, 2005, \textbf{435}, 191\relax
\mciteBstWouldAddEndPuncttrue
\mciteSetBstMidEndSepPunct{\mcitedefaultmidpunct}
{\mcitedefaultendpunct}{\mcitedefaultseppunct}\relax
\EndOfBibitem
\bibitem[Fixman and Kovac(1973)]{ref:Fixman1973}
M.~Fixman and J.~Kovac, \emph{The Journal of Chemical Physics}, 1973,
  \textbf{58}, 1564\relax
\mciteBstWouldAddEndPuncttrue
\mciteSetBstMidEndSepPunct{\mcitedefaultmidpunct}
{\mcitedefaultendpunct}{\mcitedefaultseppunct}\relax
\EndOfBibitem
\bibitem[Marko and Siggia(1995)]{ref:Marko1995}
J.~F. Marko and E.~D. Siggia, \emph{Macromolecules}, 1995, \textbf{28},
  8759--8770\relax
\mciteBstWouldAddEndPuncttrue
\mciteSetBstMidEndSepPunct{\mcitedefaultmidpunct}
{\mcitedefaultendpunct}{\mcitedefaultseppunct}\relax
\EndOfBibitem
\bibitem[Broedersz \emph{et~al.}(2012)Broedersz, Sheinman, and
  MacKintosh]{ref:Chase_length}
C.~P. Broedersz, M.~Sheinman and F.~C. MacKintosh, \emph{Phys. Rev. Lett.},
  2012, \textbf{108}, 078102\relax
\mciteBstWouldAddEndPuncttrue
\mciteSetBstMidEndSepPunct{\mcitedefaultmidpunct}
{\mcitedefaultendpunct}{\mcitedefaultseppunct}\relax
\EndOfBibitem
\bibitem[Odijk(1995)]{odijk1995stiff}
T.~Odijk, \emph{Macromolecules}, 1995, \textbf{28}, 7016--7018\relax
\mciteBstWouldAddEndPuncttrue
\mciteSetBstMidEndSepPunct{\mcitedefaultmidpunct}
{\mcitedefaultendpunct}{\mcitedefaultseppunct}\relax
\EndOfBibitem
\bibitem[Huisman \emph{et~al.}(2008)Huisman, Storm, and
  Barkema]{ref:HuismanPRE}
E.~M. Huisman, C.~Storm and G.~T. Barkema, \emph{Phys. Rev. E}, 2008,
  \textbf{78}, 051801\relax
\mciteBstWouldAddEndPuncttrue
\mciteSetBstMidEndSepPunct{\mcitedefaultmidpunct}
{\mcitedefaultendpunct}{\mcitedefaultseppunct}\relax
\EndOfBibitem
\bibitem[van Buul \emph{et~al.}(2013)van Buul, Schwartz, Brocorens, Koepf,
  Beljonne, Maan, Christianen, Kouwer, Nolte, Engelkamp, Blank, and
  Rowan]{ref:single}
A.~M. van Buul, E.~Schwartz, P.~Brocorens, M.~Koepf, D.~Beljonne, J.~C. Maan,
  P.~C.~M. Christianen, P.~H.~J. Kouwer, R.~J.~M. Nolte, H.~Engelkamp, K.~Blank
  and A.~E. Rowan, \emph{Chemical Science}, 2013, \textbf{4}, 2357\relax
\mciteBstWouldAddEndPuncttrue
\mciteSetBstMidEndSepPunct{\mcitedefaultmidpunct}
{\mcitedefaultendpunct}{\mcitedefaultseppunct}\relax
\EndOfBibitem
\bibitem[Lees and Edwards(1972)]{ref:Lees_Edwards}
A.~W. Lees and S.~F. Edwards, \emph{J. Phys. C}, 1972, \textbf{5}, 1921\relax
\mciteBstWouldAddEndPuncttrue
\mciteSetBstMidEndSepPunct{\mcitedefaultmidpunct}
{\mcitedefaultendpunct}{\mcitedefaultseppunct}\relax
\EndOfBibitem
\bibitem[Squire \emph{et~al.}(1969)Squire, Holt, and Hoover]{ref:Hoover}
D.~R. Squire, A.~C. Holt and W.~G. Hoover, \emph{Physica}, 1969, \textbf{42},
  388\relax
\mciteBstWouldAddEndPuncttrue
\mciteSetBstMidEndSepPunct{\mcitedefaultmidpunct}
{\mcitedefaultendpunct}{\mcitedefaultseppunct}\relax
\EndOfBibitem
\end{mcitethebibliography}

\end{document}